\pdfoutput=1
\documentclass[conference,compsoc]{IEEEtran}
\IEEEoverridecommandlockouts
\usepackage{cite}
\usepackage{amsmath,amssymb}
\usepackage{graphicx}
\usepackage{booktabs}
\usepackage{xcolor}
\usepackage{url}
\usepackage{eso-pic}
\usepackage[hidelinks]{hyperref}
\hypersetup{
  pdftitle={TwinGuard-Lite: A Rule-Based State-Admission Gateway for Generative Patient Digital Twins},
  pdfauthor={Wenhui Chu and Sheikh Rabiul Islam},
  pdfkeywords={Agentic AI security, data provenance, patient digital twins, prompt injection, state integrity}
}
\newcommand{\system}{\textsc{TwinGuard-Lite}}
\newcommand{\zt}{z_{n,t}}

\newcommand{\NCand}{29{,}131\pm2706}
\newcommand{\NMal}{1090\pm16}
\newcommand{\KwRecall}{0.16}
\newcommand{\AnomRecall}{0.24}
\newcommand{\AnomF}{0.190\pm0.017}
\newcommand{\AnomFPR}{0.049\pm0.006}
\newcommand{\ProvRecall}{0.599\pm0.009}
\newcommand{\PoneRecall}{0.605\pm0.009}
\newcommand{\PthreeRecall}{0.300\pm0.005}
\newcommand{\DsixRecall}{0.905\pm0.008}
\newcommand{\DsixPrec}{0.991\pm0.006}
\newcommand{\DsixF}{0.946\pm0.005}
\newcommand{\DsixFPRpct}{0.033\%}
\newcommand{\KwRecallPct}{16\%}
\newcommand{\AnomRecallPct}{24\%}
\newcommand{\ProvRecallPct}{59.9\%}
\newcommand{\DsixPrecPct}{99.1\%}
\newcommand{\DsixRecallPct}{90.5\%}
\newcommand{\DsixFPct}{94.6\%}
\newcommand{\DsixFPcount}{9.35\pm5.75}
\newcommand{\NBenign}{28{,}041}
\newcommand{\FPRAllBenignPct}{0.033\%}
\newcommand{\NRuleApplicable}{2997}
\newcommand{\FPRApplicablePct}{0.31\%}
\newcommand{\NConflictNamed}{66}
\newcommand{\FPRConflictNamedPct}{14.2\%}
\newcommand{\NThinStays}{3}
\newcommand{\NThinMin}{15}
\newcommand{\NTreatment}{12{,}172}
\newcommand{\TreatmentPct}{12\%}
\newcommand{\AtkPrevalence}{3.7\%}
\newcommand{\NPairDrugNamed}{28.5}
\newcommand{\FPRPairDrugNamedPct}{32.8\%}
\newcommand{\NPartnerOpportunities}{187}
\newcommand{\NRaw}{103{,}821}
\newcommand{\NRawMed}{16{,}335}
\newcommand{\NDropped}{6137}
\newcommand{\NDroppedPct}{38\%}
\newcommand{\FPTotal}{187}
\newcommand{\FPRule}{187}
\newcommand{\FPPatients}{8}
\newcommand{\InjResidual}{24\%}
\newcommand{\PonebTP}{145}
\newcommand{\PonebTPx}{1}
\newcommand{\PonebSpoofTP}{144}
\newcommand{\NSeeds}{20}
\newcommand{\NPeople}{362}
\newcommand{\NStays}{500}
\newcommand{\NCalibPeople}{253}
\newcommand{\NTestPeople}{109}
\newcommand{\NTestStaysMean}{148\pm7}
\newcommand{\NEpisodes}{424}
\newcommand{\NTestEpisodes}{133}
\newcommand{\FamInj}{436}
\newcommand{\FamRetr}{328}
\newcommand{\FamXpat}{327}
\newcommand{\FPPerSeed}{9.35}
\newcommand{\FPNoAttack}{187}
\newcommand{\FPAttackCaused}{0}
\newcommand{\FPPair}{amiodarone/digoxin}
\newcommand{\FPNPairs}{1}
\newcommand{\JointTP}{19{,}742}
\newcommand{\OverlapTP}{1}
\newcommand{\SpoofTotal}{2211}
\newcommand{\SpoofTargeted}{171}
\newcommand{\SpoofTargetablePct}{7.7\%}
\newcommand{\PonebRecallTargeted}{84\%}
\newcommand{\PonebRecallSpoof}{6.5\%}
\newcommand{\KwLoud}{1.00}
\newcommand{\KwQuiet}{0.00}
\newcommand{\KwSpoof}{0.00}
\newcommand{\AnomLoud}{0.84}
\newcommand{\AnomQuiet}{0.19}
\newcommand{\AnomSpoof}{0.17}
\newcommand{\AnomRetr}{0.16}
\newcommand{\AnomXpat}{0.04}
\newcommand{\ConfPrec}{0.764\pm0.141}
\newcommand{\ConfRecall}{0.028\pm0.008}
\newcommand{\ConfFPcount}{9.40\pm5.78}
\newcommand{\ConfFPTotal}{188}
\newcommand{\SensEpisodes}{500}
\newcommand{\SensDsixRecall}{0.904\pm0.008}
\newcommand{\SensDsixPrec}{0.993\pm0.004}
\newcommand{\SensDsixFPcount}{7.25\pm4.25}
\newcommand{\SensInjResidual}{24\%}
\newcommand{\SensPonebTP}{134}
\newcommand{\RetrSpoofSuccess}{92.9\%}
\newcommand{\RetrSpoofRecall}{0.626\pm0.010}
\newcommand{\RetrSpoofF}{0.766\pm0.008}
\newcommand{\RetrSpoofFPTotal}{187}

\begin{document}

\title{\textsc{TwinGuard-Lite}: A Rule-Based State-Admission Gateway\\
for Generative Patient Digital Twins}

\author{\IEEEauthorblockN{Wenhui Chu and Sheikh Rabiul Islam}
\IEEEauthorblockA{Department of Cybersecurity\\
College of Emergency Preparedness, Homeland Security and Cybersecurity\\
University at Albany, State University of New York\\
Albany, NY, USA\\
wchu3@albany.edu, sislam7@albany.edu}}

\AddToShipoutPictureBG*{\AtPageUpperLeft{\put(\LenToUnit{0.75in},\LenToUnit{-0.42in}){%
\parbox[t]{\dimexpr\paperwidth-1.5in\relax}{\scriptsize\centering
\copyright~2026 IEEE. Personal use of this material is permitted. Permission from IEEE must be obtained for all other uses, in any current or future media, including reprinting/republishing this material for advertising or promotional purposes, creating new collective works, for resale or redistribution to servers or lists, or reuse of any copyrighted component of this work in other works.}}}}
\maketitle

\begin{abstract}
Future generative patient digital twins may combine longitudinal health records
with language-model agents and keep information across sessions. The wording of
a proposed update does not reveal whether it comes from an allowed source,
conflicts with the patient's record, or belongs to someone else. We present
\system, a rule-based gateway that admits an update to a twin's persistent state
only if it passes checkable approximations of two integrity properties. Grounded
state consistency (GSC) is approximated by a provenance check and a contradiction
check; cross-patient noninterference (CPN) is approximated by a namespace check
against trusted transport metadata. We evaluate three attack
families over $\NSeeds$ seeded person-level splits of a semi-synthetic stream
built from the openly licensed eICU database demo, with $\NCand$ candidate
updates per seed. A keyword filter and an anomaly detector detect \KwRecallPct{}
and \AnomRecallPct{} of attacks, respectively; a provenance-only ablation detects
\ProvRecallPct{} but misses every cross-patient attack. Combining GSC and CPN yields \DsixPrecPct{} precision,
\DsixRecallPct{} recall, a \DsixFPct{} $F_1$ score, and a \DsixFPRpct{}
false-positive rate. Every attack the full gateway admits falsely claims a
trusted source; when the retrieval channel does so, \RetrSpoofSuccess{} of
retrieval attacks are admitted. The results hold under stated trust
assumptions, and no language model, agent, or retriever is executed: \system{}
is a mechanism-level proof of concept that motivates authenticated,
patient-bound ingestion, not a clinical safeguard.
\end{abstract}

\begin{IEEEkeywords}
Agentic AI security, data provenance, patient digital twins, prompt injection,
state integrity
\end{IEEEkeywords}

\section{Introduction}
Patient digital twins---patient-specific digital replicas that inform clinical
decisions---are still largely preclinical, mostly simulation or monitoring
models~\cite{katsoulakis2024digital,drummond2024patient}. We consider a future
\emph{generative} twin that keeps state across sessions and acts through
language-model agents and clinical tools. Such a twin inherits the security
problems of agentic AI: indirect prompt
injection~\cite{zhan2024injecagent,debenedetti2024agentdojo}, retrieval
poisoning~\cite{zou2025poisonedrag}, memory
poisoning~\cite{chen2024agentpoison}, and poisoning of medical
models~\cite{alber2025poisoning}. Recent HealthSec work traces the same attack
paths through a healthcare system powered by large language models
(LLMs)~\cite{nagaraja2025goal}. The
patient setting adds a distinctive asset: a \emph{persistent, identity-bound
clinical state}. A corrupted fact survives across sessions and, if patient
namespaces are not isolated, can spread between people.

Content filters and anomaly detectors are tempting first defenses. The attacks
that evade them, however, are \emph{textually unremarkable}: a fact phrased like a
legitimate record but lacking authorized provenance, a claim that silently
contradicts an existing record, or a value copied from another patient. Such a
write is invalid because of its source, its relation to the record, or its
owner, not because of its wording. Following design-level prompt-injection
defense~\cite{debenedetti2026camel} and classical integrity
policy~\cite{clark1987comparison,goguen1982noninterference}, we therefore enforce
patient-twin integrity through explicit rules checked on each state update.

We make three contributions. First, we define four integrity properties, one
for each of four relations a state update has (Section~\ref{sec:model}). Second,
we build \system, a rule-based gateway that enforces checkable approximations of
two of them before any update persists (Section~\ref{sec:proto}). Third, we
evaluate it on a semi-synthetic, real-data-backed benchmark with three attack
families, two surface baselines, and a check-by-check ablation
(Section~\ref{sec:eval}). Section~\ref{sec:limits} discusses limits.

\section{Model and Integrity Properties}
\label{sec:model}

\subsection{Twin State and Candidate Writes}
A twin for patient namespace $n$ at logical time $t$ holds a canonical state
$\zt$ of accepted clinical facts, a persistent memory, a capability set
$\kappa_{n,t}$, and an append-only event and provenance ledger. A patient is a
person, not an encounter. No input changes $\zt$ or $\kappa_{n,t}$ directly:
every input first becomes a \emph{candidate write}, a proposed update that
carries an event type, a claimed namespace, an origin label, an episode (one
hospital admission), an event time, and a value. Inputs come from three trust zones (Fig.~\ref{fig:arch}):
\emph{authoritative patient evidence} (authenticated but possibly erroneous),
\emph{curated knowledge} (versioned reference material, including the retrieval
store, not evidence about this patient), and \emph{untrusted context} (free text,
unvetted external content, model output). Only authoritative evidence may write patient state.

\subsection{Integrity Properties}
We consider four relations a write stands in: to its source and the record it
joins, to its owner, to the sequence of states before it, and to the actions it
enables. Asking one question about each relation defines one property.

\emph{Is it supported?} \textbf{Grounded state consistency (GSC):} every
accepted fact $f\in\zt$ has an unbroken provenance path to authoritative
evidence and violates no declared contradiction rule, unless an explicit
unresolved-conflict state records the violation.

\emph{Whose is it?} \textbf{Cross-patient noninterference (CPN):} for patients
$i\neq j$, changing only untrusted inputs in $i$'s history cannot change the
canonical state, memory, or action trace of $j$, apart from a declared,
minimized allow-list of shared metadata that carries no patient-distinguishing
signal.

\emph{Does it follow from the history?} \textbf{Longitudinal transition
integrity (LTI):} the sequence of admitted states follows a declared transition
relation. \emph{What may it trigger?} \textbf{Authorized action integrity (AAI):} every
action taken on twin state is permitted by an explicit action policy.

The four properties mirror classical integrity policy. GSC corresponds to the
Clark--Wilson rule that unconstrained inputs are either validated into
constrained data or rejected~\cite{clark1987comparison}. LTI corresponds to the
Clark--Wilson requirement that transformations move data from one valid state
to another, and AAI to its rules on who may run which transformation. CPN is
noninterference~\cite{goguen1982noninterference} with patients as the security
domains and the allow-list as the only sanctioned downgrading channel.

The properties do not collapse into one another: individually grounded writes
can trace an illegal trajectory (LTI), a grounded write can land in the wrong
record (CPN), admissible state can drive an unauthorized action (AAI), and a
correctly attributed write can be unsourced (GSC). We do not claim
completeness; we argue coverage against \mbox{Clark--Wilson} instead. The append-only
ledger realizes its logging rule, while periodic verification of stored data and
separation of duty are not modeled. The set targets the integrity of persistent
state against an attacker who can propose writes. Legality of transitions over
time falls under LTI; interval-aware contradiction rules remain GSC checks on the
current record. Clinical correctness of authoritative evidence is left to
clinical validation, and confidentiality, availability, and model integrity are
out of scope. \system{} enforces per-write approximations of GSC and CPN, the
two properties that need no transition relation or action policy. Because CPN
relates two runs, the namespace check enforces a stronger per-write condition:
no write crosses namespaces outside the allow-list. Under a trusted origin
binding, this approximates CPN for direct writes on the canonical-state channel,
the only channel implemented. Influence routed through shared stores, such as a
retrieval index, needs lineage tracking: the namespace check does not see it,
and only the GSC checks can stop it.

\subsection{Threat Model and Trust Assumptions}
The adversary seeks a target-state change, cross-patient influence, or an
unauthorized action, and enters through a prompt, a retrieved document, a memory
candidate, or a tool output. We parameterize capability rather than assume one
universal attacker~\cite{vassilev2025aml}. The evaluated attacker proposes
candidate writes, may spoof the mutable origin label, and knows the frozen
two-pair contradiction policy (Section~\ref{sec:bench}), but it is nonadaptive
at run time. It cannot
alter the policy engine or the \emph{true} originating namespace, which the
benchmark supplies as trusted transport metadata.

Two assumptions bound every result. (i) CPN is evaluated \emph{given} a correct
binding between each write and the patient it came from. A deployment must
obtain that binding from an authenticated session or a signed, patient-bound
event envelope. Our experiment validates the policy under this binding but does
not show how to establish the binding; without the binding, CPN inherits the
label-spoofing weakness of the provenance check (Section~\ref{sec:proto}). (ii) Evaluation starts once a poisoned item
has become a candidate write. We model neither index construction nor retrieval
ranking, so we study \emph{retrieval-mediated state poisoning} rather than
end-to-end retrieval-augmented generation (RAG) poisoning.

\section{\system: A Rule-Based Gateway}
\label{sec:proto}

\begin{figure}[!t]
\centering
\includegraphics[width=\columnwidth]{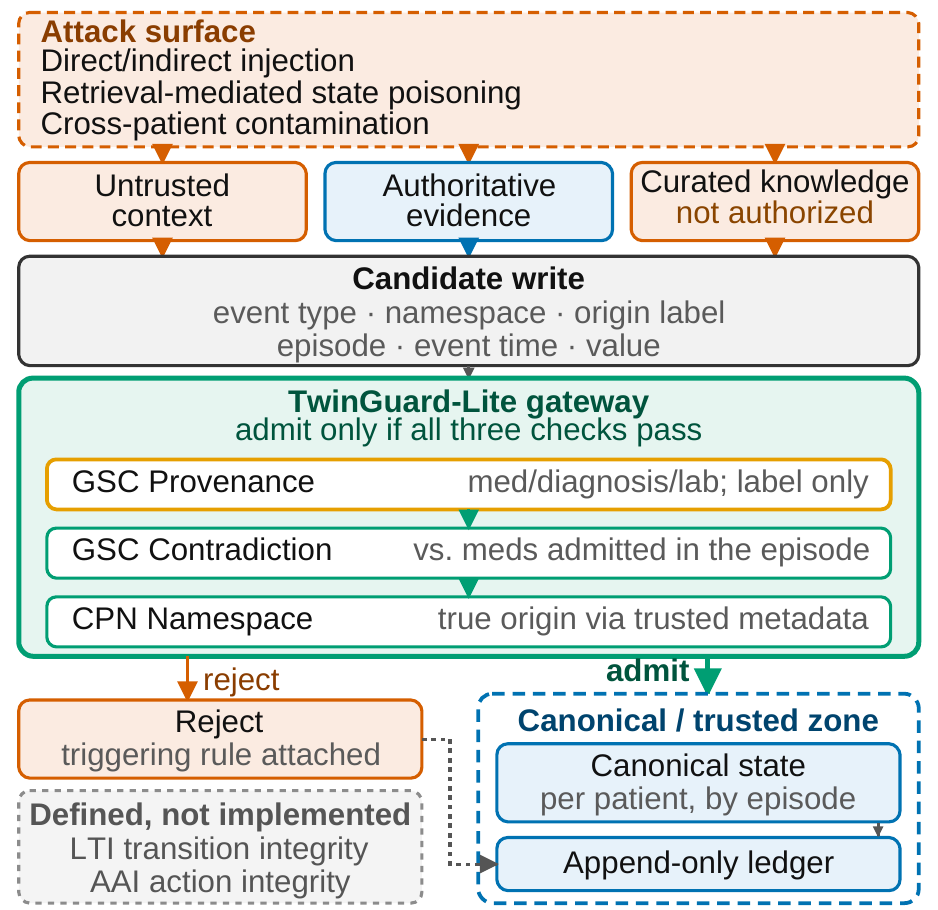}
\caption{Architecture of \system. The gateway
applies the provenance and contradiction checks (GSC) and the namespace check
(CPN); a write is admitted only if all three pass, and every decision is logged.
Orange marks unauthorized origins and rejection, amber the spoofable
label-only check, blue the authoritative-evidence zone and the canonical/trusted zone,
green the gateway and admission, and gray the candidate write and what is
defined but not implemented. A provenance spoof forges an authoritative label; a
cross-patient write carries an origin valid for its source patient but claims
the target's namespace; poisoned retrieval enters through the curated-knowledge
zone.}
\label{fig:arch}
\end{figure}

Candidate writes from all three trust zones go through the three checks of
Fig.~\ref{fig:arch}, described here from top to bottom.

(1) The \emph{provenance check} (GSC) rejects a medication, diagnosis, or
laboratory candidate whose origin label is not authoritative. It reads only the
label: it builds no provenance chain and verifies no signature, so it is cheap
but spoofable. Treatment candidates are outside its scope.

(2) The \emph{contradiction check} (GSC) rejects a medication candidate that
forms a declared conflict pair with a medication already admitted to the same
episode; drug names are matched after normalization. The prototype implements no
unresolved-conflict state, so every match is rejected.

(3) The \emph{namespace check} (CPN) rejects a candidate whose true originating
namespace, taken from trusted transport metadata, differs from the namespace it
claims, unless its type is on the frozen shared-metadata allow-list.

\emph{How the checks combine.} Admission is a conjunction: a write enters
canonical state only if all three checks pass, and any single failure rejects
it. The checks are side-effect-free predicates, so their order decides only
which rule the ledger records as the reason, never whether a write is admitted.
We chose conjunction because each check encodes a hard invariant, not a score;
weighted or learned fusion would let strong evidence on one property buy
admission for a violation of another. Beyond the ablation arms and surface
baselines in Table~\ref{tab:det}, we have not evaluated other combination rules;
risk-scored routing of borderline writes to clinician review is future work. Every decision is appended to the ledger
together with its triggering rule, if any. The prototype keeps admitted medication state
only for the contradiction check and implements no full clinical twin state,
memory store, capability system, or action trace.

\begin{table*}[!t]
\centering
\caption{Detection performance over $\NSeeds$ seeded person-level splits
(mean $\pm$ population standard deviation (SD); $\NCand$ candidates and $\NMal$
malicious per seed).
The lower block ablates the gateway check by check (GSC: provenance +
contradiction; CPN: namespace). The false-positive rate (FPR) is a fraction;
FP/seed counts false
positives per seed. The SD reflects resplitting one fixed cohort and redrawing
attacks, not population sampling.}
\label{tab:det}
\footnotesize
\begin{tabular}{lccccc}
\toprule
Defense & Precision & Recall & $F_1$ & FPR & FP/seed \\
\midrule
Generator-matched keyword filter & $1.000\pm0.000$ & $0.158\pm0.011$ & $0.273\pm0.016$ & $0.00000\pm0.00000$ & $0.00\pm0.00$ \\
Anomaly detector (2$\sigma$) & $0.160\pm0.023$ & $0.236\pm0.011$ & $0.190\pm0.017$ & $0.04919\pm0.00583$ & $1385.65\pm256.08$ \\
\midrule
Provenance only & $1.000\pm0.000$ & $0.599\pm0.009$ & $0.749\pm0.007$ & $0.00000\pm0.00000$ & $0.00\pm0.00$ \\
Contradiction only & $0.764\pm0.141$ & $0.028\pm0.008$ & $0.054\pm0.016$ & $0.00034\pm0.00021$ & $9.40\pm5.78$ \\
GSC only (provenance + contradiction) & $0.986\pm0.009$ & $0.605\pm0.009$ & $0.750\pm0.008$ & $0.00033\pm0.00021$ & $9.35\pm5.75$ \\
CPN only (namespace) & $1.000\pm0.000$ & $0.300\pm0.005$ & $0.461\pm0.006$ & $0.00000\pm0.00000$ & $0.00\pm0.00$ \\
\textbf{TwinGuard-Lite (GSC+CPN)} & $\mathbf{0.991\pm0.006}$ & $\mathbf{0.905\pm0.008}$ & $\mathbf{0.946\pm0.005}$ & $\mathbf{0.00033\pm0.00021}$ & $\mathbf{9.35\pm5.75}$ \\
\bottomrule
\end{tabular}

\end{table*}

\section{Preliminary Evaluation}
\label{sec:eval}

\subsection{Benchmark}
\label{sec:bench}
Benign trajectories come from the eICU Collaborative Research Database
(\mbox{eICU-CRD}) Demo v2.0.1~\cite{johnson2021eicudemo}, the openly licensed
subset of the full \mbox{eICU-CRD}~\cite{pollard2018eicu}. Namespaces are people,
not intensive care unit (ICU) stays: our
$\NStays$ unit stays resolve to $\NPeople$ people, and we split by person,
$\NCalibPeople$ for calibration and $\NTestPeople$ ($\NTestStaysMean$~stays) for
testing, so no person lies on both sides. Times in eICU are offsets from each
unit stay's own admission. We therefore rebuild a shared \emph{episode} clock
for each hospital admission as \texttt{offset}~$-$~\texttt{hospitalAdmitOffset}
and replay admitted medication state within each episode ($\NEpisodes$ episodes;
$\NTestEpisodes$ in the first seed's test split). Events cover medications, diagnoses, laboratory results,
and treatments.

The cohort is the $\NStays$ smallest \texttt{patientunitstayid} values among
Demo stays with at least $20$ extracted events. This rule is exactly
reproducible but not random: stay identifiers are not assigned at random with
respect to hospital, and the event threshold favors data-rich stays. The cohort
spans $48$ distinct de-identified values in the source \texttt{hospitalid} field
(dataset codes, not a verified count of real sites), so the results should not
be read as generalizing across sites or acuities. Once missing payloads are
removed, $\NThinStays$ stays fall below $20$ usable events ($\NThinMin$ at the
minimum); we keep them rather than reselect. Drug names are
missing for $\NDropped$ of the $\NRawMed$ medication rows ($\NDroppedPct$;
$\NRaw$ rows overall); they are dropped at load, so no medication attack or
vocabulary uses a missing-value marker. However, $559$ retained laboratory rows lack a numeric result,
and cross-patient copies of them are still rejected by the namespace check. The medication table's coded fields
(\texttt{drugHiclSeqno}, \texttt{GTC}) are not used.

Only the benign stream is observed data; every adversarial write, origin label,
and namespace binding is generated by the benchmark. Three attack families are
injected into the test split, giving $\NCand$ candidates per seed,
\mbox{$\NMal$} of them malicious. \emph{Direct/indirect injection} has \emph{loud} (instruction
tokens), \emph{quiet} (a plain fact), and \emph{provenance-spoof} (a forged
authoritative label) subtypes. Like the spoofs, \emph{retrieval-mediated state
poisoning} targets the declared conflict pairs, but it carries a
\texttt{curated} origin. \emph{Cross-patient contamination} copies an
event type and its exact value from a \emph{different person's} real stream.
The spoof attacker is \emph{targeted} and \emph{causal}: it reads only earlier
records in the episode it writes to and places a real surface form of a
declared-conflict drug at a strictly later episode time. The contradiction rule
matches each declared drug name as a whole word anywhere in the payload, because
eICU text often opens with a package quantity
(\texttt{2 ML - DIGOXIN 0.25 MG/ML IJ SOLN}). It covers only string-matchable
conflicts and, in this corpus, one of the two declared pairs:
apixaban never appears, so warfarin/apixaban never fires and every
contradiction decision comes from amiodarone/digoxin.

\subsection{Defenses and Protocol}
The surface baselines are a keyword filter over the loud-injection generator's
own instruction tokens and an anomaly detector. The detector scores each write by its
per-character negative log-likelihood under a Laplace-smoothed fourth-order
character Markov model fitted on calibration text only, and flags scores at
least two standard deviations ($2\sigma$) above the mean score of benign
calibration writes. Two choices keep this baseline honest. Text is case-folded and
whitespace-collapsed before scoring, and every injected medication value is a
verbatim corpus string; otherwise the detector would separate typography, not
content. The structured arms ablate the gateway check by check: provenance
only, contradiction only, GSC only (provenance + contradiction), CPN only
(namespace), and \system{} (GSC+CPN). A no-defense arm flags nothing and gives
the attack-success baseline.

We report precision, recall, $F_1$, false-positive rate (FPR), and per-family
attack success (the fraction of malicious writes admitted to canonical state).
Each metric is computed per seed and macro-averaged, as mean $\pm$ population
standard deviation (SD) over $\NSeeds$ seeds. This spread reflects resplitting
one fixed cohort and redrawing attacks, not uncertainty about the eICU
population or independent sites, and a macro-averaged ratio may differ slightly
from the same ratio recomputed from pooled counts. Per-subtype and
spoof-targeting rates are instead pooled over seeds, as are the false-positive
ratios in Section~\ref{sec:results}, a pooled attribution audit dividing the
$\FPTotal$ false-positive decisions by four nested benign counts. Given a seed,
the pipeline is deterministic, and same-minute writes are ordered only by fields
the gateway observes.

\subsection{Results}
\label{sec:results}
\begin{figure}[!t]
\centering
\includegraphics[width=\columnwidth]{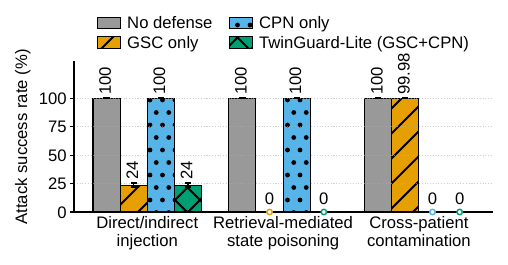}
\caption{Attack success by family (lower is better; mean $\pm$ population SD
over $\NSeeds$ seeds): the percentage of each family's malicious writes admitted
to canonical state, out of $\FamInj$/$\FamRetr$/$\FamXpat$
injection/retrieval/cross-patient attempts per seed (rounded means).
Table~\ref{tab:det} gives detection metrics for all arms. An open marker is a
measured $0$; $99.98\%$ is printed to two decimals to distinguish it from the
definitional $100\%$. The $0\%$ and $100\%$ values follow from the policy
construction and trust assumptions (Sections~\ref{sec:model}
and~\ref{sec:limits}), not from a general detector.}
\label{fig:succ}
\end{figure}

Table~\ref{tab:det} and Fig.~\ref{fig:succ} summarize the main run. The surface
baselines are weak for a structural reason. The keyword filter recalls only
$\KwRecall$: it catches every loud injection ($\KwLoud$), while quiet, spoofed,
poisoned, and cross-patient writes carry no token to match ($\KwQuiet$ on quiet
injection, $\KwSpoof$ on the provenance spoofs). The anomaly detector recalls
$\AnomRecall$ ($F_1=\AnomF$). It too sees mainly loud writes ($\AnomLoud$),
falling to $\AnomQuiet$, $\AnomSpoof$, $\AnomRetr$, and $\AnomXpat$ on quiet,
spoofed, retrieval-mediated, and cross-patient writes, respectively. It also
flags $\AnomFPR$ of the mean $\NBenign$ benign writes per seed, which rules it out as a standalone
admission control. Provenance alone reaches recall $\ProvRecall$ with no false
positives but misses every cross-patient write, whose origin is authoritative
for its source patient. GSC alone reaches $\PoneRecall$ and CPN alone
$\PthreeRecall$. Only their combination reaches recall $\DsixRecall$ (precision
$\DsixPrec$, $F_1=\DsixF$) at a \DsixFPRpct{} FPR, or $\DsixFPcount$ rejected
benign writes per seed, and it drives retrieval-mediated poisoning and, given
the trusted binding, cross-patient contamination to $0\%$ success.

\emph{Why the combined row is nearly a sum.} The recall and false-positive
count of \system{} are almost exactly the sums of the GSC-only and CPN-only
values. This is not because one check is selected per write; every write must
pass all three. The two properties catch nearly disjoint attacks, sharing only
$\OverlapTP$ of $\JointTP$ pooled true positives (largely by construction;
Section~\ref{sec:limits}), and CPN adds no false positives.

\emph{The contradiction check alone.} By itself, the contradiction check reaches
only recall $\ConfRecall$ (precision $\ConfPrec$; $\ConfFPcount$ false positives
per seed), because it inspects only medication writes that name a declared
drug. Its pooled false positives rise from $\FPTotal$ (GSC only) to
$\ConfFPTotal$ because a malicious medication write that the provenance check
would have rejected is admitted and becomes a conflict partner of a later benign
write. It is therefore not a detector in its own right, but it is the only check
that can stop a write with a forged authoritative label. In the full gateway,
$\PonebTP$ rejections are logged under it, all provenance spoofs except
$\PonebTPx$ cross-patient write that the namespace check also rejects. It thus
stops $\PonebSpoofTP$ attacks that no other check stops but causes all
$\FPTotal$ benign rejections, more than it alone prevents; its worth therefore
depends on a better-sourced rule set.

That role matters because the $0\%$ retrieval result reflects trust-zone
enforcement, not robustness to poisoned retrieval: the provenance check rejects
the family's \texttt{curated} origin first. We reran all seeds with the
retrieval channel \emph{lying} about its origin (\texttt{authoritative}), so
that only the contradiction check can still reject those writes. Then
$\RetrSpoofSuccess$ of that family is admitted, and \system{} falls to recall
$\RetrSpoofRecall$ ($F_1=\RetrSpoofF$), with the benign stream and its
$\RetrSpoofFPTotal$ false positives unchanged. The residual $\InjResidual$
injection success in the main run likewise consists entirely of provenance
spoofs. Only
$\SpoofTargeted$ of $\SpoofTotal$ spoofs ($\SpoofTargetablePct$) could be
targeted at all---the rest landed in episodes with no drug whose declared
partner the attacker could supply---and the check caught $\PonebRecallTargeted$ of those, or
$\PonebRecallSpoof$ of all spoofs. Contradiction checking is a partial backstop
against spoofing, not a substitute for authenticated ingestion. With the strictest
episode boundary, where no state crosses a unit stay ($\SensEpisodes$
episodes), the conclusions hold: recall
$\SensDsixRecall$, precision $\SensDsixPrec$, injection residual
$\SensInjResidual$, $\SensDsixFPcount$ false positives per seed, and
$\SensPonebTP$ pooled contradiction true positives.

\emph{Where the false positives come from.} All $\FPTotal$ pooled
false-positive decisions came from the contradiction rule ($\FPRule$ of
$\FPTotal$) on $\FPNPairs$ pair (\FPPair), over $\FPPatients$ people whose
episodes contain both drugs ($\FPPerSeed$ per seed; repeated splits retest the
same records, so these are decisions, not distinct records). A matched
no-attack counterfactual---the same benign stream, order, and gateway, with
every malicious write deleted---yields $\FPNoAttack$ rejections and
$\FPAttackCaused$ changed decisions, so in the full gateway these false
positives are the cost of the rule itself, not collateral damage from attacks. The whole-stream FPR
understates this cost, since most benign writes never meet a contradiction rule.
Pooled, the same $\FPTotal$ decisions are $\FPRAllBenignPct$ of all benign
writes (as in Table~\ref{tab:det}) and $\FPRApplicablePct$ of the
$\NRuleApplicable$ benign medication writes per seed the rule is scoped to. They
are $\FPRConflictNamedPct$ of the $\NConflictNamed$ per seed naming one of the
four drugs with a declared partner and $\FPRPairDrugNamedPct$ of the
$\NPairDrugNamed$ naming a drug from a \emph{corpus-supported} pair. The
$\FPRAllBenignPct$ rate is what a deployment would see; the
$\FPRPairDrugNamedPct$ rate characterizes the mechanism. The partner must also
be admitted earlier in the episode, which happens $\NPartnerOpportunities$ times
in the pooled benign stream, equal to the numerator by construction because
every match is rejected. Precision also depends on prevalence:
attacks are $\AtkPrevalence$ of writes here by construction, and at a realistic,
much lower prevalence the same FPR would dominate precision.

\section{Discussion and Limitations}
\label{sec:limits}
We discuss the main limits, including those raised in review, and our view of
each.

\emph{Strength by construction.} The benign data is real, but the origin
labels, namespace bindings, and attacks are generated, so the structured checks
are strong partly by construction. Likewise, the keyword list is the loud
generator's own, so its $\KwLoud$ loud recall is guaranteed, and neither
baseline is state of the art. The near-disjointness of GSC and CPN detections
also follows from the family construction, and the injection and retrieval
families violate the same provenance check, so their separate rates report
attack provenance, not independent defenses.

\emph{Trusted patient binding.} CPN's $0\%$ cross-patient success holds
\emph{given} a trusted true-origin binding. Once that binding exists, detecting a
crossing is simple; the hard problem is establishing the binding through
authenticated sessions or signed, patient-bound event envelopes, and this work
does not solve it. The CPN
allow-list is also unvalidated and unexercised: no evaluated write uses an
allow-listed type, so the CPN numbers are those of a deny-every-crossing rule.

\emph{Authenticated provenance.} The provenance check reads a label, and the
spoofed-retrieval run shows the consequence: $\RetrSpoofSuccess$ of lying
retrieval writes are admitted. Authenticated, origin-bound provenance is
therefore central to making the approach usable, and the label check stands in
for it here. The check also covers only medication, diagnosis, and laboratory
writes, so the $\NTreatment$ treatment rows ($\TreatmentPct$ of the corpus) pass
it unexamined. No attack exploits this gap: treatment-type attacks occur only as
cross-patient copies, which the namespace check rejects.

\emph{Contradiction coverage.} The declared list holds two drug pairs that we
chose as demonstration policy, not drawn from an interaction authority,
guideline, or prespecified protocol. It therefore carries a cherry-picking risk
as well as a coverage limit, and the reported contradiction recall is an upper
bound conditional on the list. Our extraction omits \texttt{drugStopOffset}, so
medication state is cumulative and a flagged pair may never have been active at
the same time. Even an active pair need not be an error: amiodarone with
digoxin, the pair behind every false positive here, can be prescribed
deliberately with a reduced digoxin dose. A larger, clinically grounded rule
set with medication start and stop times would make the evaluation more
convincing and reduce such rejections. Until then, flagged writes should be
\emph{quarantined for clinician adjudication}, with the ledger entry and rule
attached, rather than silently rejected.

\emph{End-to-end evaluation.} No language model, agent, or retriever runs here.
Starting at the admission boundary isolates the mechanism: the decision depends
only on the candidate write, its transport metadata, and the admitted state,
though the producing pipeline sets the fields the checks read, notably the
origin label. It does not show how often, or in what form, real injected or
retrieved content becomes a candidate write. A small end-to-end study, in which an LLM agent
with retrieval produces writes that reach the gateway, is the most important
next experiment. The attacker is also nonadaptive at run time. Ongoing work
adds authenticated ingestion, a maintained rule set with medication intervals,
LTI and AAI enforcement, adaptive attackers, and patient-clustered statistics.

\section{Related Work}
Agentic prompt-injection benchmarks and
defenses~\cite{debenedetti2024agentdojo,zhan2024injecagent,debenedetti2026camel,evtimov2025wasp},
retrieval and memory poisoning~\cite{zou2025poisonedrag,chen2024agentpoison},
and medical-model poisoning~\cite{alber2025poisoning} establish the individual
attack surfaces, and adversarial machine learning
taxonomies~\cite{vassilev2025aml} structure our threat model. At HealthSec'25,
Nagaraja and Bahsi~\cite{nagaraja2025goal} apply goal-driven risk assessment to
an LLM-powered healthcare system, including prompt injection and knowledge-base
poisoning, and identify cross-patient context contamination through poor memory
isolation as a risk; CPN addresses the integrity side of that risk for
persistent state. At HealthSec'24, Hallajiyan et~al.~\cite{hallajiyan2024sam}
anticipate inference-time false data injection attacks on
machine-learning-enabled medical devices. Both study how corrupted inputs reach clinical
systems; we add an enforcement point for the state they leave behind.

CaMeL~\cite{debenedetti2026camel} stops untrusted data from steering control
flow \emph{within} one execution; \system{} governs what an execution may persist
\emph{across} executions. Neither subsumes the other: a protected run can still
commit an unsourced fact, and CPN quantifies over a second patient's history,
which no single trace expresses.

\section{Conclusion}
Attacks that evade surface filters are textually ordinary but structurally
invalid. Enforcing rule-level approximations of GSC and CPN as auditable
admission checks detects \DsixRecallPct{} of the malicious writes here at a
\DsixFPRpct{} false-positive rate. Every admitted attack is a provenance spoof,
calling for authenticated ingestion, and every rejected benign write comes from
the contradiction rule, calling for rules from a maintained interaction
authority. The main practical takeaway is a workflow for systems that persist
patient state: bind every write to an authenticated source and patient, check
it against declared rules before it persists, log each decision, and quarantine
what the rules flag. As
no language model, agent, or retriever is run, these results validate defined
properties under stated trust assumptions, not clinical readiness.

\section*{Ethics and Data}
Benign streams derive from the open-access \mbox{eICU-CRD} Demo
v2.0.1~\cite{johnson2021eicudemo}, which PhysioNet~\cite{pollard2026physionet}
distributes under the Open Database License (ODbL)~1.0 without a credentialing
agreement; it is a de-identified subset of the full
\mbox{eICU-CRD}~\cite{pollard2018eicu}. \mbox{Privacert} certified its re-identification risk as meeting the Health Insurance Portability and
Accountability Act (HIPAA) Safe Harbor standard (Certification no.~1031219-2).
The study involves no interaction with human subjects and no direct
identifiers, and we make no
attempt to re-identify anyone; our derived tables remain under ODbL~1.0 with
attribution. All adversarial writes are benchmark-generated; no real eICU record
is labeled malicious.

\section*{Acknowledgments}
We are deeply grateful to the three anonymous reviewers of \mbox{HealthSec'26}
for their careful, generous, and constructive reviews, which substantially
improved this paper. They asked how the integrity properties were chosen and
whether they are complete, how the implemented properties combine, and what the
contradiction check contributes on its own. They helped us examine the trust
assumptions behind the patient binding and the provenance check, the scope of
the contradiction rules, and the missing end-to-end experiment, and they showed
us how to organize the paper and name its properties clearly. The paper is
clearer and more precise about its assumptions because of them. We also thank the \mbox{HealthSec'26} workshop chair and
program committee for their guidance.

\bibliographystyle{IEEEtran}
\bibliography{refs}

\begin{thebibliography}{10}
\providecommand{\url}[1]{#1}
\csname url@samestyle\endcsname
\providecommand{\newblock}{\relax}
\providecommand{\bibinfo}[2]{#2}
\providecommand{\BIBentrySTDinterwordspacing}{\spaceskip=0pt\relax}
\providecommand{\BIBentryALTinterwordstretchfactor}{4}
\providecommand{\BIBentryALTinterwordspacing}{\spaceskip=\fontdimen2\font plus
\BIBentryALTinterwordstretchfactor\fontdimen3\font minus
  \fontdimen4\font\relax}
\providecommand{\BIBforeignlanguage}[2]{{%
\expandafter\ifx\csname l@#1\endcsname\relax
\typeout{** WARNING: IEEEtran.bst: No hyphenation pattern has been}%
\typeout{** loaded for the language `#1'. Using the pattern for}%
\typeout{** the default language instead.}%
\else
\language=\csname l@#1\endcsname
\fi
#2}}
\providecommand{\BIBdecl}{\relax}
\BIBdecl

\bibitem{katsoulakis2024digital}
E.~Katsoulakis \emph{et~al.}, ``Digital twins for health: A scoping review,''
  \emph{npj Digital Medicine}, vol.~7, no.~1, p.~77, 2024.

\bibitem{drummond2024patient}
D.~Drummond and A.~Gonsard, ``Definitions and characteristics of patient
  digital twins being developed for clinical use: Scoping review,''
  \emph{Journal of Medical Internet Research}, vol.~26, p. e58504, 2024.

\bibitem{zhan2024injecagent}
Q.~Zhan, Z.~Liang, Z.~Ying, and D.~Kang, ``{InjecAgent}: Benchmarking indirect
  prompt injections in tool-integrated large language model agents,'' in
  \emph{Findings of the Association for Computational Linguistics: ACL 2024},
  2024, pp. 10\,471--10\,506.

\bibitem{debenedetti2024agentdojo}
E.~Debenedetti, J.~Zhang, M.~Balunovi{\'c}, L.~Beurer-Kellner, M.~Fischer, and
  F.~Tram{\`e}r, ``{AgentDojo}: A dynamic environment to evaluate prompt
  injection attacks and defenses for {LLM} agents,'' in \emph{Advances in
  Neural Information Processing Systems, Datasets and Benchmarks Track},
  vol.~37, 2024.

\bibitem{zou2025poisonedrag}
W.~Zou, R.~Geng, B.~Wang, and J.~Jia, ``{PoisonedRAG}: Knowledge corruption
  attacks to retrieval-augmented generation of large language models,'' in
  \emph{34th USENIX Security Symposium (USENIX Security 25)}, 2025, pp.
  3827--3844.

\bibitem{chen2024agentpoison}
Z.~Chen, Z.~Xiang, C.~Xiao, D.~Song, and B.~Li, ``{AgentPoison}: Red-teaming
  {LLM} agents via poisoning memory or knowledge bases,'' in \emph{Advances in
  Neural Information Processing Systems}, vol.~37, 2024.

\bibitem{alber2025poisoning}
D.~A. Alber \emph{et~al.}, ``Medical large language models are vulnerable to
  data-poisoning attacks,'' \emph{Nature Medicine}, vol.~31, no.~2, pp.
  618--626, 2025.

\bibitem{nagaraja2025goal}
N.~Nagaraja and H.~Bahsi, ``Goal-driven risk assessment for {LLM}-powered
  systems: A healthcare case study,'' in \emph{Annual Computer Security
  Applications Conference Workshops (ACSACW), HealthSec'25}, 2025, pp.
  290--297.

\bibitem{debenedetti2026camel}
E.~Debenedetti \emph{et~al.}, ``Defeating prompt injections by design,'' in
  \emph{IEEE Conference on Secure and Trustworthy Machine Learning (SaTML)},
  2026.

\bibitem{clark1987comparison}
D.~D. Clark and D.~R. Wilson, ``A comparison of commercial and military
  computer security policies,'' in \emph{IEEE Symposium on Security and
  Privacy}, 1987, pp. 184--194.

\bibitem{goguen1982noninterference}
J.~A. Goguen and J.~Meseguer, ``Security policies and security models,'' in
  \emph{IEEE Symposium on Security and Privacy}, 1982, pp. 11--20.

\bibitem{vassilev2025aml}
A.~Vassilev, A.~Oprea, A.~Fordyce, H.~Anderson, X.~Davies, and M.~Hamin,
  ``Adversarial machine learning: A taxonomy and terminology of attacks and
  mitigations,'' NIST, Tech. Rep. AI 100-2e2025, 2025.

\bibitem{johnson2021eicudemo}
\BIBentryALTinterwordspacing
A.~Johnson, T.~Pollard, O.~Badawi, and J.~Raffa, ``{eICU Collaborative Research
  Database Demo} (version 2.0.1),'' PhysioNet, 2021. [Online]. Available:
  \url{https://doi.org/10.13026/4mxk-na84}
\BIBentrySTDinterwordspacing

\bibitem{pollard2018eicu}
T.~J. Pollard, A.~E.~W. Johnson, J.~D. Raffa, L.~A. Celi, R.~G. Mark, and
  O.~Badawi, ``The {eICU Collaborative Research Database}, a freely available
  multi-center database for critical care research,'' \emph{Scientific Data},
  vol.~5, p. 180178, 2018.

\bibitem{evtimov2025wasp}
I.~Evtimov, A.~Zharmagambetov, A.~Grattafiori, C.~Guo, and K.~Chaudhuri,
  ``{WASP}: Benchmarking web agent security against prompt injection attacks,''
  in \emph{Advances in Neural Information Processing Systems, Datasets and
  Benchmarks Track}, vol.~38, 2025.

\bibitem{hallajiyan2024sam}
M.~Hallajiyan, A.~P. Dharmalingam, G.~Mitra, H.~Alemzadeh, S.~Iqbal, and
  K.~Pattabiraman, ``{SAM}: Foreseeing inference-time false data injection
  attacks on {ML}-enabled medical devices,'' in \emph{Workshop on Cybersecurity
  in Healthcare (HealthSec'24)}, 2024, pp. 77--84.

\bibitem{pollard2026physionet}
T.~Pollard \emph{et~al.}, ``{PhysioNet} as a global platform for biomedical
  research,'' \emph{Nature Health}, vol.~1, no.~8, pp. 792--795, 2026.

\end{thebibliography}

\end{document}